\documentstyle[preprint,aps]{revtex}
\begin{document}
\draft
\preprint{\vbox{Submitted to Physical Review C\hfill
        UC/NPL-1125}}
\tolerance = 10000
\hfuzz=5pt
\tighten
\begin{title}
{The electron-nucleon cross section in $(e,e'p)$ reactions}
\end{title}
\author{S. Pollock}
\address{
Department of Physics, CB 390,\\
University of Colorado, Boulder CO 80309-0446\\
Email address: pollock@lucky.colorado.edu}
\author{H.W.L. Naus}
\address{
Institute for Theoretical Physics,\\
University of Hannover}
\author{J.H. Koch}
\address{
National Institute for Nuclear
Physics and High Energy Physics (NIKHEF), Amsterdam\\
and\\
Institute for Theoretical Physics, University of Amsterdam\\
}
\maketitle

\begin{abstract}
We examine commonly used approaches to deal with the scattering of
electrons from a bound nucleon. Several prescriptions are shown to be
related by gauge transformations. Nevertheless, due to current
non-conservation, they yield different results. These differences
reflect the size of the uncertainty that persists in the interpretation
of $(e,e'p)$ experiments.
\end{abstract}
\section{Introduction}

In the interpretation of electron-nucleus scattering experiments one
must make a choice of how to describe the interaction between an
electron and a bound nucleon. Only the scattering of an electron on a
free, on-shell nucleon is determined model independently. The
kinematics of the scattering on a bound, off-shell nucleon is
necessarily different and therefore there exists no well defined unique
procedure for the theoretical description of the nuclear scattering
process.

In trying to describe the nuclear reaction by means of the free
electromagnetic current of the nucleon, assumptions have to be made.
They lead to a non-conserved nuclear current, an unphysical feature
that is usually remedied in an {\it ad hoc} fashion.  The most commonly
used `conserved current' (cc) prescription for the $(e,e'p)$ reaction
was introduced by de Forest \cite{kn:def83}.  This prescription also
makes it possible to factorize the PWIA cross section into a part
containing the electron-nucleon cross section and a nuclear structure
part.  By comparing some variations within this class of recipes, it is
often concluded that the uncertainty due to this procedure is small and
that 'off-shell' effects are negligible.

Clearly, this last point needs to be critically examined before one can
draw conclusions from {\it e.g.} $(e,e'p)$ experiments about subtle or
exotic effects, either concerning nuclear structure or the influence of
the medium on the reaction mechanism. An example of a reaction where
this consideration enters is the recent $(e,e'p)$ measurement by
Makins {\it et al.} \cite{kn:mak94}.  It was motivated by the
suggestion of a particular medium effect, color transparency.

It is the purpose of this note to briefly review the various
approximations which go into the standard descriptions of the $(e,e'p)$
reaction and result in a non-conserved nuclear current. We discuss in
detail prescriptions to restore conservation of the electromagnetic
current of the off-shell nucleon and relate them to particular choices
of a gauge.  Since there is much interest in the $(e,e'p)$ experiment
by Makins {\it et al.} \cite{kn:mak94}, we give examples for the
kinematics of this experiment even though they are at the peak of the
quasielastic cross section and the initial nucleon is not far off its
mass shell. Our general conclusion is that the ambiguities connected
to the electromagnetic current of an off-shell nucleon cannot be
dismissed even if predictions among some currently used prescriptions
are in close agreement.

\section{Current conserving prescriptions}

There has been considerable work on general aspects of the
electromagnetic interaction with the nucleons in a nucleus (see {\it e.g.}
\cite{kn:fru84}, \cite{kn:na90}, \cite{kn:ito91},
\cite{kn:gro92},
\cite{kn:hen92},
\cite{kn:so92}).
The nuclear wavefunction, the electromagnetic vertex
and {\it e.g.} the
final state interaction need to be dealt with consistently. We will not
repeat this discussion here and comment only on the assumptions that go
into the often used recipe by de Forest \cite{kn:def83} for the cross
section for a bound, off mass shell nucleon. They are good examples for
the problems one encounters in general and for the approximations one
makes in practice.

The general form of the nuclear current is
\begin{equation}
J_{\mu} = \Psi_f \Gamma_{\mu} \Psi_i,
\end{equation}
where $\Psi_{i,f}$ denote the initial and final wavefunctions and
$\Gamma_\mu$ is the electromagnetic vertex operator.  It is quite
common to consider only the contributions due to one body currents. In
practice, to obtain a manageable description additional {\it ad hoc}
assumptions are made concerning the wavefunctions, the vertex operator,
the kinematics and current conservation.  For simplicity, we will
consider the $(e,e'p)$ reaction in PWIA, where the initial nucleon is
bound and the final one is in a plane wave, on mass shell state.

{\it Wavefunction:} The assumption made in Ref. \cite{kn:def83} is that
the wavefunction of both the plane wave final nucleon and also the
initial bound nucleon is given by the Dirac spinor for an on-shell
nucleon. For the initial nucleon it is assumed that this spinor is
determined through its three momentum, $\vec{p}$, the missing momentum
of the initial nucleon, and the corresponding on-shell energy,
$E_{on}=\sqrt{\vec{p}^{\,2}+M^2}$.

{\it Vertex operator:} The general vertex for an off-shell nucleon,
appearing between the nucleon wavefunctions, has been discussed in the
literature, {\it e.g.} in Ref. \cite{kn:bi60}.  The operator structure
can be much more complex than the one one encounters in expressions for
the free current. Furthermore, the associated form factors can depend
in addition to $q^{2}$, the photon four-momentum, on other scalar
variables such as  the invariant mass of the initial nucleon, $p^{2}$.
Rather than using this general expression (which would prevent
factorization), all commonly used recipes make use of the free
current.  However, there are a variety of ways to write the free
on-shell current in terms of two independent vertex operators and
associated form factors.  De Forest uses two forms
\begin{equation}
J^{\mu}_{1}=e\bar{u}(\vec{p}')\left\{ [F_1(q^2)+
                              F_2(q^2)]\gamma^\mu
       -F_2(q^2)\frac{(p+p')^\mu}{2M}\right\}u(\vec{p}),
\label{eq:cur1}
\end{equation}
and
\begin{equation}
J^{\mu}_{2}=e\bar{u}({\vec{p}\,}')\left\{F_{1}(q^{2})\gamma^{\mu}
+F_{2}(q^{2})\frac{i\sigma^{\mu\nu}q_{\nu}}{2M}\right\}u(\vec{p}),
\label{eq:cur2}
\end{equation}
which can be transformed into each other by means of the Gordon
decomposition. While for on-shell nucleons the two currents are
equivalent, the results obtained when one tries to use them in the
off-shell case are different.

{\it Kinematics:} In the $(e,e'p)$ reaction the energy transfer by the
electron, $\omega$, and the energy of the detected nucleon, $E'$,
determine the energy of the initial bound nucleon to be $E=E'-\omega
\neq E_{on}$.  However, the use of a free on-shell spinor in the
construction of the current involves the on-shell energy $E _{on}$ for
the initial nucleon. In the current based on eq. (\ref{eq:cur1}), the
energy of the initial nucleon also appears explicitly not only in the
spinor, but also in the vertex operator and the usual prescription is
to use $E _{on}$ in the operator. An alternative is discussed in Ref.
\cite{kn:na90}.

{\it Current conservation:} After the above manipulations, it is clear
that the resulting current is not conserved. The last step then is to
make the current conserved by hand. We will discuss three possibilities
to do this and apply these methods to the two ways to write the free
on-shell current, eqs. (\ref{eq:cur1}) and (\ref{eq:cur2}).

(a) The method chosen in Ref. \cite{kn:def83} is to replace the longitudinal
component $J _{q}$, parallel to $\vec{q}$, by the charge
density $J_{0}$:
\begin{equation}
J_{q} \rightarrow J_{q}'= \frac{\omega J_{0}}{|\vec{q}\,|}.
\label{eq:jl}
\end{equation}
and thus work with a four-current
\begin{equation}
J_{\mu}^{( \ell )} = (\vec{J}_{t}, \frac{\omega J_{0}}{|\vec{q}\,|}, J_{0}).
\label{eq:Jl}
\end{equation}
This would be correct and of no consequence if the current indeed was
conserved. It has been argued that Siegerts theorem suggests this
substitution  when the current is not exactly conserved, but this long
wavelength argument doesn't apply for the one-body current one is
concerned with here, nor can it be expected to hold at the energies we
consider below. The cross sections arising from this recipe, the often
used prescriptions by de Forest, will be referred to simply as
`$\sigma_{cc}$'.

(b) Of course, one could take care of current conservation in
the opposite way by eliminating the charge density instead \cite{kn:na90},
\cite{kn:cab90}:
\begin{equation}
J_{0} \rightarrow J_{0}'= \frac{\vec{J} \cdot \vec{q}}{\omega},
\label{eq:j0}
\end{equation}
and to use
\begin{equation}
J_{\mu}^{(0)} =(\vec{J}, \frac{\vec{J} \cdot \vec{q}}{\omega}).
\label{eq:j0p}
\end{equation}
The resulting cross section will be referred to as $\sigma_{cc}^{0}$.

(c) In other recipes \cite{kn:mou76}
one subtracts a term proportional to $q_{\mu}$ to obtain a divergence free
current:
\begin{equation}
J_{\mu} \rightarrow J_{\mu}^{(q)}= J_{\mu}-\frac{J \cdot q}{q^2}q_{\mu}.
\label{eq:js}
\end{equation}
The cross section obtained from this recipe will be referred to as
$\sigma_{cc}^{q}$.

{\it Connection to the gauge choice:} As will be shown below, these
different ways to restore current conservation can be seen as a choice
of a gauge, which in principle should have no effect on the results.
That these choices lead to different results shows the inconsistencies
inherent in the commonly chosen approach to deal with the
electromagnetic interaction of bound nucleons.
The electron scattering matrix element can be written as
\begin{equation}
M=j^{\mu} \Pi_{\mu\nu} J^{\nu},
\label{eq:mael}
\end{equation}
where $\Pi$ denotes the photon propagator and $j$ the electron current.
The explicit form of the propagator is gauge dependent and, as a
consequence, so is the form of the matrix element.

In the covariant Lorentz class of gauges one has
\begin{equation}
M_{L}=\frac{i}{q^2}(-j \cdot J +(1-\xi)\frac{(q \cdot J)(q \cdot j)}{q^2}),
\label{eq:Lor}
\end{equation}
where $\xi$ is a free gauge parameter.
It is common practice to work in the Feynman gauge, $\xi=1$. In this case,
one obtains
\begin{equation}
M_{F}=\frac{i}{q^2}(-j \cdot J ).
\label{eq:Fey}
\end{equation}
This of course is always the case in the covariant Lorentz gauges since the
electron current, $j$, is conserved and the second term in Eq.
(\ref{eq:Lor}) vanishes. We will now show that the matrix elements
resulting from the above three modified `conserved' currents, eqs.
(\ref{eq:Jl}), (\ref{eq:j0p}), and (\ref{eq:js}), when used in the
Feynman gauge yield the same matrix elements one obtains with the
original, non-conserved current, but evaluated in different gauges.

{\it Coulomb gauge:} The well-known Coulomb gauge is an example of a
non-covariant gauge. Using the Coulomb gauge propagator for
$\Pi_{\mu\nu}$, the general matrix element, eq. (\ref{eq:mael}),
reduces to
\begin{equation}
M_{C}=\frac{i}{\vec{q}^{\,2}}j_{0}J_{0}+\frac{i}{q^2}(\vec{j} \cdot \vec{J}
-\frac{(\vec{q} \cdot \vec{J})(\vec{q} \cdot \vec{j})}{\vec{q}^{\,2}}).
\label{eq:Cou}
\end{equation}
This is precisely the same matrix element one would obtain in the Feynman
gauge, upon using the replacement given in eq. (\ref{eq:jl}).
The second part of eq. (\ref{eq:Cou}) is the contribution of the
transverse parts of the current,
defined as
\begin{equation}
\vec{J}_{t}=\vec{J}-\frac{\vec{q} \cdot \vec{J}}{\vec{q}^{\,2}}\vec{q}.
\label{eq:jt}
\end{equation}
Depending on whether one uses the current $J^{\mu}_1$ given in eq.
(\ref{eq:cur2}) or $J^{\mu}_2$ eq. (\ref{eq:cur1}), one obtains
$\sigma_{cc1}$ and $\sigma_{cc2}$ from $M_{C}$. These are the widely
used cross sections proposed by de Forest \cite{kn:def83}.

{\it Weyl gauge:} Another non-covariant gauge is the Weyl (or temporal)
gauge.  Using the photon propagator in this gauge, the charge densities
do not explicitly contribute to the matrix element:
\begin{equation}
M_{W}=\frac{i}{q^2}(\vec{j} \cdot \vec{J}
-\frac{(\vec{q} \cdot \vec{J})(\vec{q} \cdot \vec{j})}{\omega^{2}}).
\label{eq:Weyl}
\end{equation}
Again, it is readily seen that this is the same expression one would
have obtained in the Feynman gauge upon using the replacement given in
eq.  (\ref{eq:j0}), yielding $\sigma_{cc1}^{0}$ or $\sigma_{cc2}^{0}$,
depending on the form for the on-shell current one used to approximate
the off-shell current.

{\it Landau gauge:} Finally, another example from the covariant Lorentz
class is the Landau gauge, defined by the gauge parameter $\xi = 0$.
As one can see from eq. (\ref{eq:Lor}), this yields $\sigma_{cc1}^q$
and $\sigma_{cc2}^q$, the same result as in the Feynman gauge with the
{\it ad hoc} subtraction defined in eq. (\ref{eq:js}) that guarantees a
conserved current. In fact, one would obtain this result if one did
nothing and simply used the original non-conserved current in eq.
(\ref{eq:Fey}).

Of course, physical observables should not depend on the choice of the
gauge. Indeed, for conserved currents all the matrix elements given
above can easily be shown to be equivalent. However, for non-conserved
currents, {\it i.e.} broken gauge invariance, choosing a different gauge
gives a different result. This is the situation for the approximation
for the bound nucleon current: the results are not the same. The choice
of which component to eliminate in favor of another or to simply make
the {\it ad hoc} subtraction, eq. (\ref{eq:js}), can thus be related
to the choice of a gauge. The connection between a choice of the gauge
and non-contributing parts of the currents is formally always present.
However, it is only exact for conserved currents.

\section{Results and Conclusions}

{\it Estimates of the differences between cc prescriptions:} The formal
connection between gauge choices and different cc prescriptions can be
used for getting estimates of the uncertainties {\it within} the
cc-class.  The starting point is that the nucleon current $J$, is not
conserved.  Different matrix elements are obtained in non-covariant
gauges. Since the electron current is conserved, all covariant Lorentz
class gauges yield the same result. These differences between the cc
recipes will be used below for different kinematics to get an
impression of the uncertainty introduced by dealing with the off-shell
current in an {\it ad hoc} fashion. It should be emphasized that the
differences can only give a rough indication of these ambiguities as a
function of the relevant kinematical variables. These estimates are not
based on any dynamical input, but only on the connection between the cc
prescriptions explained in the previous section.

A measure of how far one is from the on-shell kinematics is provided by
the energy transfer. The actual energy transfer to the nucleon,
$\omega$, is determined by the electron kinematics. If the initial
nucleon was on its mass shell, its energy $E_{on}$ would be
$(\vec{p}^{\,2} + M^2)^{1/2}$,
where $\vec{p}$ is the missing momentum. The energy transfer, $\omega'$, which
one would have in that case is given by
\begin{equation}
\omega ' = E' - E_{on} .
\label{eq:omp}
\end{equation}
How far one is off-shell is therefore indicated by the difference,
$\Delta\omega$,
\begin{equation}
\Delta\omega=\omega - \omega ' .
\label{eq:delom}
\end{equation}

In Figs. 1 through 4 we show results for the off-shell electron-nucleon
scattering cross section for the various cc choices. We choose
kinematics which correspond roughly to the extremes of the kinematics
sampled by Makins {\it et al.} \cite{kn:mak94}. Shown are the
deviations of different prescriptions from $\sigma_{cc2}$, the
prescription used in Ref. \cite{kn:mak94} for the interpretation of
their data.  The cross sections are plotted as a function of $\gamma$,
the angle \cite{kn:def83} between the outgoing proton and the direction
of $\vec q$. Positive $\gamma$ corresponds to protons scattered {\it
between} the incident beam direction and $\vec q$, negative $\gamma$ is
for protons scattered beyond $\vec q$. (The experimental data in Ref.
\cite{kn:mak94} correspond to negative $\gamma$ only.) All the figures
assume that the recoil proton is in the electron scattering plane.
Note that as $|\gamma|$ increases, the missing momentum generally also
increases. We have chosen ranges of $\gamma$ which correspond to
missing momentum up to $\approx$~250 MeV.

The electron scattering kinematics in Fig. 1 is $Q^2 = 1.04$ GeV$^2$,
$|\vec q\,| = 1.2$ GeV, and the cross sections are shown for $|\vec
p\,'| = |\vec q\,|$, {\it i.e.} in perpendicular kinematics. The
missing energy is 47 MeV at the center of the plot, and depends very
weakly on $\gamma$. ($E_m=45$ MeV at $\gamma=\pm 12^\circ$) The missing
momentum ranges from 0 to 250 MeV$/c$, resulting in a $\Delta\omega$
from 47 to 80 MeV. The curves correspond to different prescriptions:
how the current is made to be conserved (or which gauge is chosen) and
which on-shell form for the current is used to start with, eq.
(\ref{eq:cur1}) or (\ref{eq:cur2}).  We see that there is a spread of
more than $\pm 5 \%$ among the different prescriptions relative to
$\sigma_{cc2}$.

In Fig. 2, we fix the momentum of the knocked out nucleon at a value
{\it lower} than $|\vec q\,|$, in order to access a larger missing
energy. In this case, with $|\vec p\,'|$ reduced by 10\% from its value
in Fig. 1, the missing energy is approximately~140 MeV at $\gamma=0$,
and the missing momentum ranges from 120 to 270 MeV$/c$.  This leads to
an increased $\Delta\omega$ between 148 and 180 MeV.  Consequently, the
largest difference between the cross sections grows to more than $\pm
10\%$.

In Fig. 3, we use the kinematics of the measurement with the highest
incident energy: $Q^2 = 6.8$ GeV$^2$, $|\vec q\,| = 4.5\ $GeV, again in
perpendicular kinematics  with $|\vec p\,'| = |\vec q\,|$; the missing
energy is 9 MeV at $\gamma$=0. In this case
one is closer to the on-shell kinematics: $\Delta\omega$ is between 9
and 40 MeV and the differences between cross sections typically around
$1\%$. In Fig. 4, $|\vec p\,'|$ is reduced (by 3\%) to access a
higher missing energy and momentum. In this case the missing energy is
137 MeV at $\gamma=0$, (135 MeV at $\gamma=\pm 3^\circ$ )
and the missing momentum ranges from 130 to 280 MeV$/c$,
resulting in a $\Delta\omega$ from 148 to  179 MeV, comparable to
Fig. 2, and the spread among the prescriptions grows to about $5\%$.

It should be stressed that variations of up to $10\%$ occur solely due
to the choice of gauge, indicating the severity of the approximations
used to make the current conserved. The figures also illustrate another
- somewhat smaller - uncertainty due to another assumption:
differences between recipes labeled as `1' and `2', {\it i.e.} show the
effect of choosing one of the two equivalent ways to write the on-shell
current as given in eqs.  (\ref{eq:cur1}) and (\ref{eq:cur2}).  For
given electron kinematics, also this difference grows as we go away
from on-shell kinematics, {\it i.e.} for larger $\Delta\omega$.

That the cross sections appear somewhat less sensitive to gauge choices
at the higher energy kinematics can be understood from the following
qualitative estimates which apply to a fixed choice of the on-shell
current.  A measure for the violation of current conservation is in
each case given by
\cite {kn:na90}
\begin{equation}
q \cdot J = \omega J_0 - \vec{q} \cdot \vec{J} \equiv \chi,\qquad
\chi \approx \Delta\omega[J],
\end{equation}
where the quantity $[J]$ denotes (part of) the nuclear current
density.
The matrix element in the Coulomb gauge, eq. (\ref{eq:Cou}), is
\begin{equation}
M_{C}=\frac{-i}{q^2}j \cdot J+ \frac{i}{q^2}(
\frac{\omega j_0 \chi }{\vec{q}^{\,2}}).
\label{eq:Cou2}
\end{equation}
Similarly, one obtains in the Weyl gauge, (\ref{eq:Weyl})
\begin{equation}
M_{W}=\frac{-i}{q^2}j \cdot J+ \frac{i}{q^2}(
\frac{j_0 \chi }{\omega}).
\label{eq:Wey2}
\end{equation}

For conserved currents, such as with the subtraction in
eq.~(\ref{eq:js}), we have $\chi=0$,  and  the matrix elements obviously
reduce to the Feynman gauge matrix expression, eq. (\ref{eq:Fey}).
Since also the electron current is conserved, the matrix elements in
all Lorentz gauges, such as Feynman and Landau gauge, are identical:
$M_F=M_L$.

With the above expressions for the matrix elements, $M_{C},M_{W}$ and
$M_L$, we can estimate the relative differences between the various
prescriptions.  We start with comparing Coulomb and Lorentz gauges.
Using eqs. (\ref{eq:Fey}) and (\ref{eq:Cou2}), one easily finds that
\begin{equation}
\frac{M_C-M_L}{M_L} \simeq -\frac{\omega j_0 \Delta\omega [J]}
{\vec{q}^{\,2} (j \cdot J)}.
\end{equation}
For the purpose of getting order of magnitude estimates, we approximate
$ j_0 [J] \simeq j \cdot J $ and find
\begin{equation}
\frac{M_C-M_L}{M_L} \simeq -\frac{\omega  \Delta\omega } {\vec{q}^{\,2} }.
\end{equation}
For a given choice of the on-shell current this expression yields the
right magnitude of the difference between the cross sections in the
figures, {\it i.e.}, the difference between
$\sigma_{cc1,2}$ and $\sigma_{cc1,2}^q$.
Similarly, one can obtain the
corresponding expression for the Weyl gauge,
\begin{equation}
\frac{M_W-M_L}{M_L} \simeq -\frac{\Delta\omega } {\omega },
\end{equation}
which gives the right magnitude for the differences between $\sigma_{cc1,2}^0$
and $\sigma_{cc1,2}^q$ .
For the comparison of Coulomb and Weyl gauges, two non-covariant gauges,
we can approximate the difference as
\begin{equation}
\frac{M_C-M_W}{M_C} \simeq \frac{-\omega  \Delta\omega
(1/\vec{q}^{\,2}-1/\omega^{2})} {1-\omega \Delta\omega/
\vec{q}^{\,2} }.
\end{equation}
In the kinematical region under consideration
this can be further approximated by
\begin{equation}
\frac{M_C-M_W}{M_C} \simeq  - \omega  \Delta\omega
(\frac{1}{\vec{q}^{\,2}}-\frac{1}{\omega^{2}}),
\end{equation}
to obtain an estimate for the differences between
$\sigma_{cc1,2}$ and $\sigma_{cc1,2}^0$. All the above estimates can
explain the relative differences among the cross sections shown in
the figures for the kinematics of the SLAC
experiment; they also explain the larger differences found in other
applications \cite{kn:na90}.

Our discussion does not provide any estimates for the differences
between prescriptions based on different on-shell currents, only for
different ways to restore current conservation.  What we have shown
are the effects due to different prescriptions in the literature for
restoring current conservation that are used in the interpretation of
$(e,e'p)$ experiments.  We also showed the variation due to
different on-shell equivalent electromagnetic currents. We have not
discussed other aspects of scattering from a bound nucleon or showed
the general framework in which all such aspects should be treated
consistently, such as the nuclear wavefunction, final state
interactions or modifications of the electromagnetic vertex operator.
The latter has been considered {\it e.g.} in meson loop models and
relatively small effects were found
\cite{kn:na87},
\cite{kn:tt}.
Until a complete and fully consistent theoretical description of the
$(e,e'p)$ reaction has been achieved, one really cannot know what a reasonable
approximation would be and which of the prescriptions we discussed is
`best'. The differences of the results we have shown give an idea of
size of the present uncertainty in the interpretation of
$(e,e'p)$ experiments.

\acknowledgments

The work of J.H.K. is part of the research program of the Foundation
for Fundamental Research of Matter (FOM) and the National Organization
for Scientific Research (NWO). S.J.P. was supported by the U.S.  D.O.E,
and acknowledges the financial support of the Sloan Foundation.

\widetext
\begin{figure}

\caption{Deviation of calculated cross
sections from de Forest's ``cc2'' prescription as a function of the
angle $\gamma$ between the ejected proton and the momentum transfer
direction.  Here incident electron energy = 2.02 GeV, $Q^2$=1.04 GeV$^2$,
$|\vec q\,|$=1.2 GeV, $|\vec{p'}\,|$=1.2 GeV, and $E_m$=47 MeV at the
center of the plot. Solid curve:
$\sigma_{cc1}$, dotted curve: $\sigma_{cc1}^0$, dashed curve:
$\sigma_{cc2}^0$, long-dashed curve: $\sigma_{cc1}^q$, dot-dashed curve:
$\sigma_{cc2}^q$.}

\label{f1}
\end{figure}
\begin{figure}
\caption{Same as Fig. 1, but with outgoing proton momentum fixed at
$|\vec{p'}\,|$=1.08 GeV, which reaches a larger missing energy
($\approx 140 MeV$ at the center of the plot.)}

\label{f2}
\end{figure}

\begin{figure}
\caption{Same as Fig. 1, but with incident energy = 5.12 GeV,
$Q^2$ = 6.77 GeV$^2$, $|\vec q\,|$=4.48 GeV,
$|\vec{p'}|\,$=4.48 GeV, and missing energy 9 MeV at $\gamma$=0.
($E_m$ = 6 MeV at $\gamma=\pm 3^\circ$)}
\label{f3}
\end{figure}

\begin{figure}
\caption{Same as Fig. 3, but with outgoing proton momentum fixed at
$|\vec{p'}\,|$=4.35 GeV, which reaches a large missing energy
($\approx 137 MeV$ at the center of the plot.)
}
\label{f4}
\end{figure}
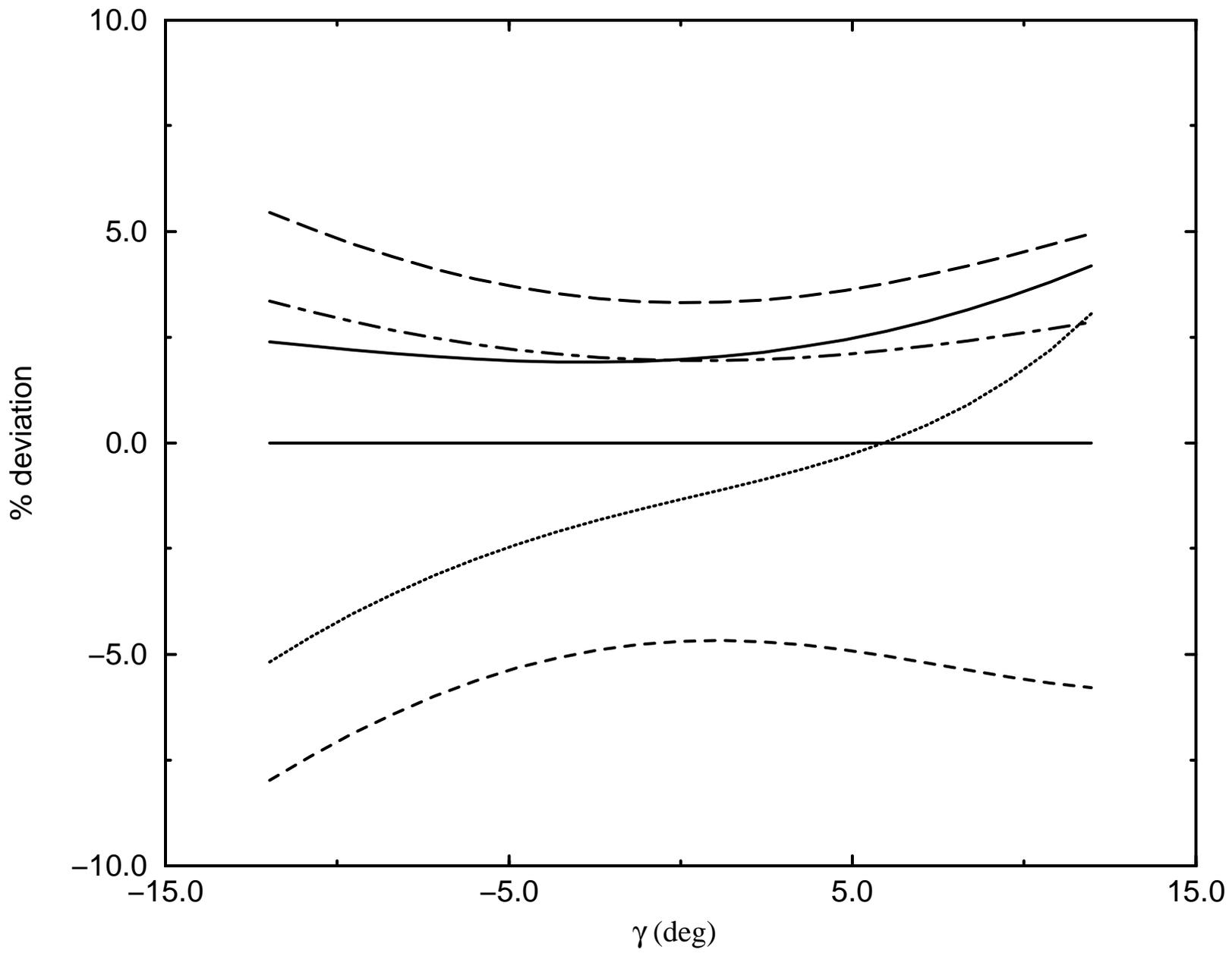
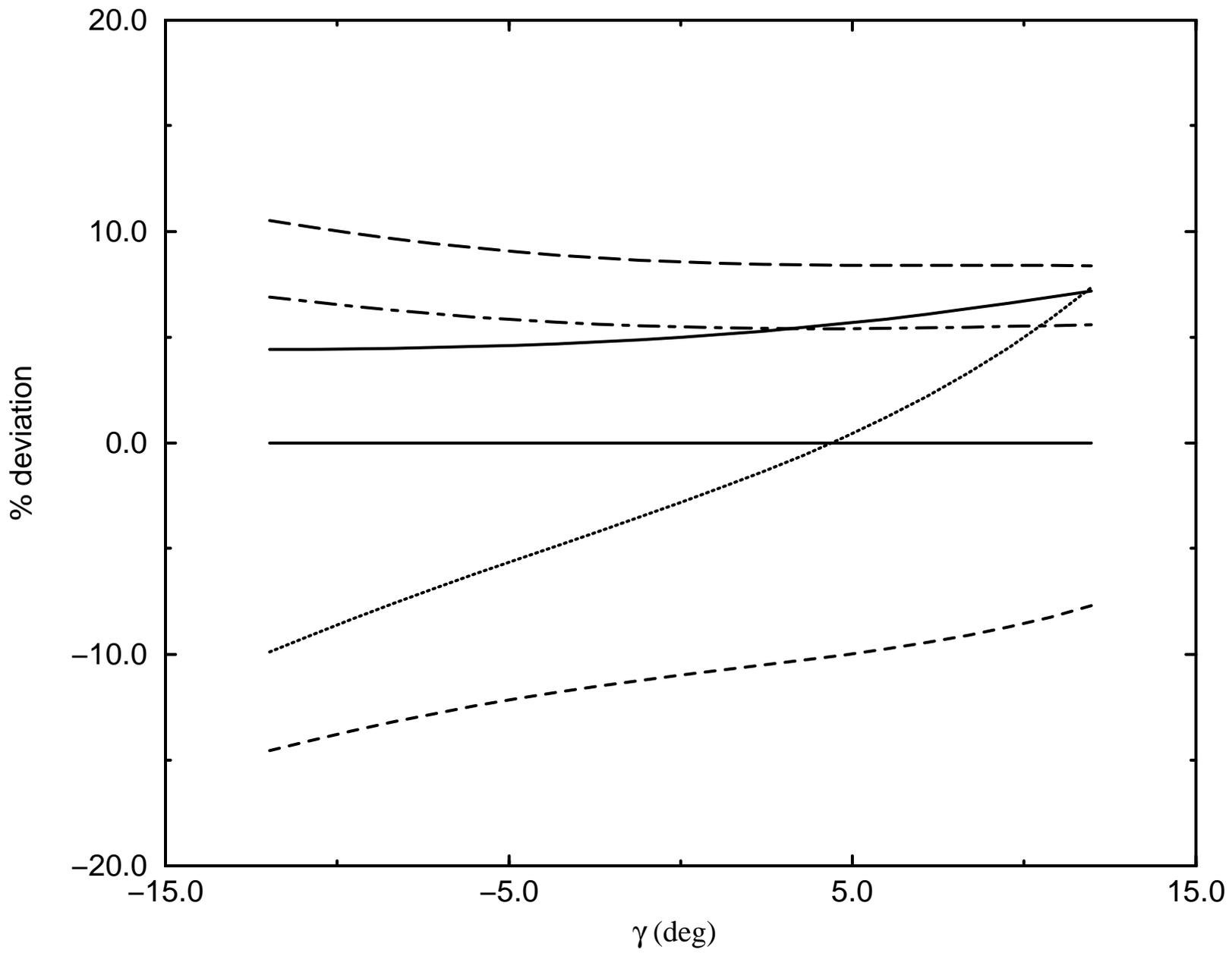
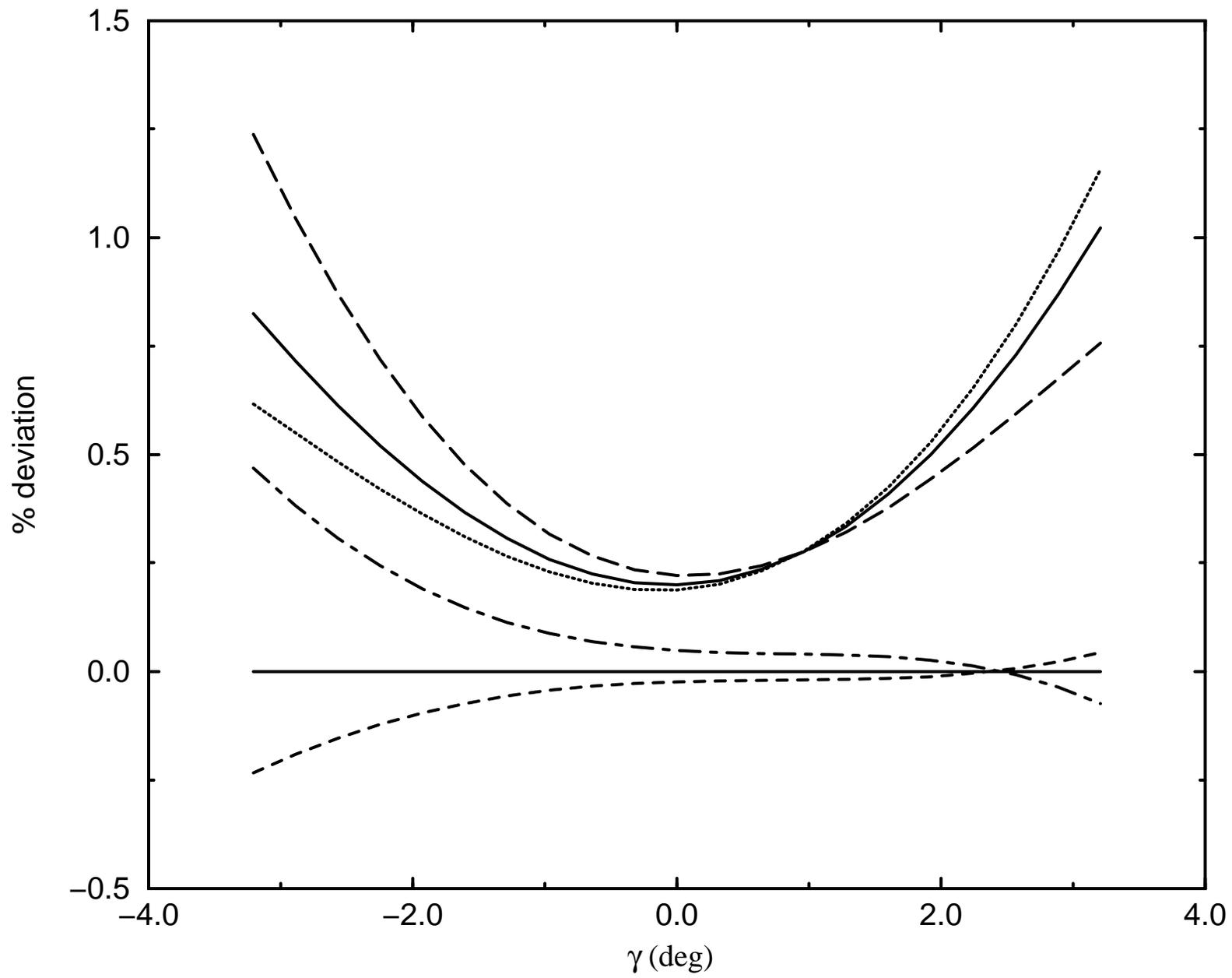
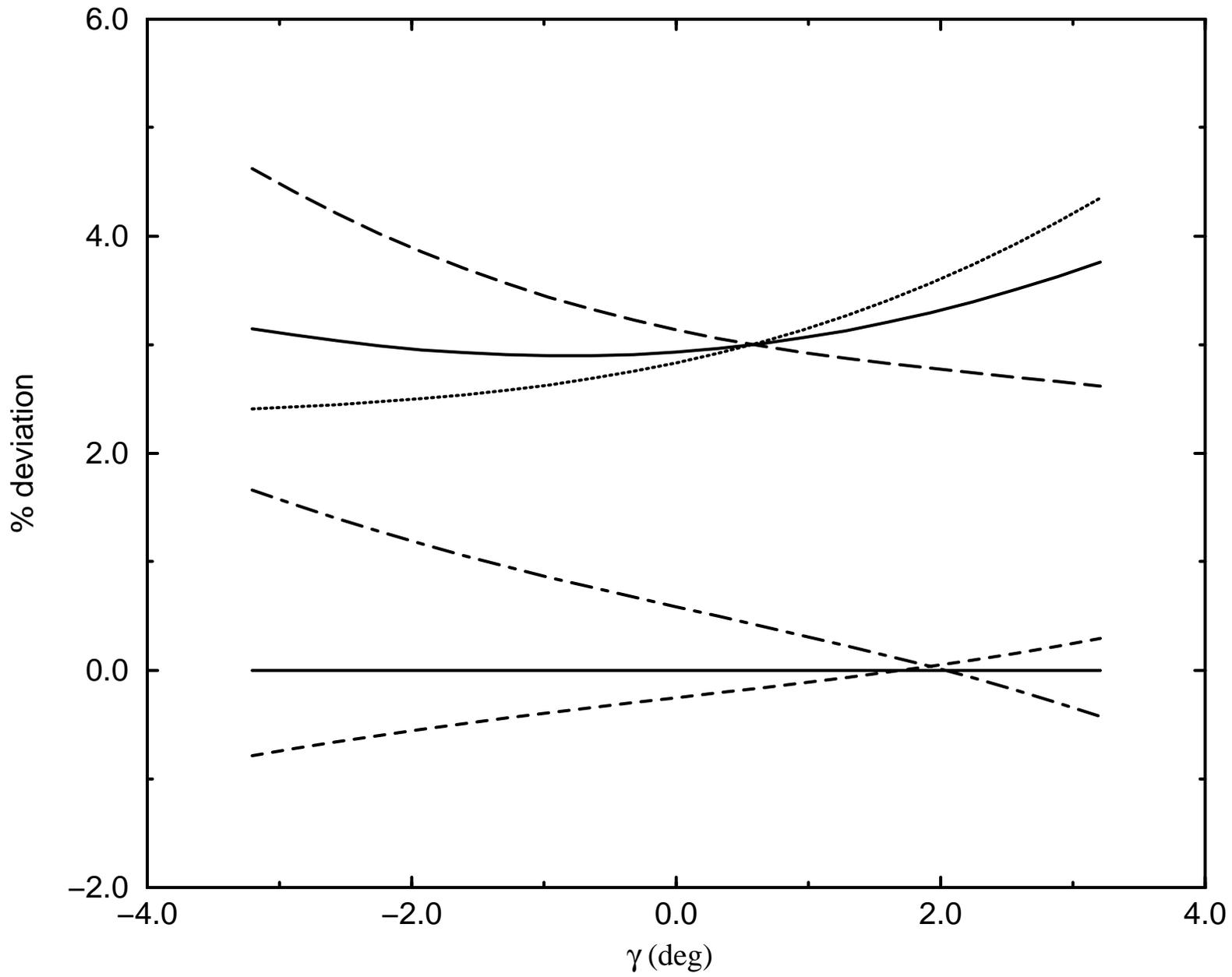
\end{document}